\documentclass{an}
\usepackage{times}
\usepackage{natbib}
\Volume{00}              % issue in current year, format {NN}
\Year{0000}              % format: {YYYY}
\Month{00}               % format: {M} or {MM} (number), month of print version
\Pagespan{000}{000}      % format: {start page}{last page}

\begin{document}
\lhead[\thepage]{A.N. S. Foucaud et al.: Observing the high redshift universe
  using the VIMOS-IFU} 
\rhead[Astron. Nachr./AN~{\bf XXX} (200X) X]{\thepage}
  \headnote{Astron. Nachr./AN {\bf 32X} (200X) X, XXX--XXX}

\title{Observing the high redshift universe using the VIMOS-IFU}

\author{S.~Foucaud\inst{1}\thanks{for the VVDS collaboration} \and
  M.~Scodeggio\inst{1}$^{\star}$ \and A.~Zanichelli\inst{2}$^{\star}$
  \and B.~Garilli\inst{1}$^{\star}$ \and E.~Giallongo\inst{3}.}
\institute{Istituto di Astrofisica Spaziale e Fisica cosmica - Sezione
  di Milano, via Bassini 15, 20133 Milano, Italy \and Istituto di
  Radioastronomia - Sezione di Bologna, via Gobetti 101, 40129
  Bologna, Italy \and Osservatorio Astronomico di Roma, via Frascati
  33, 00040 Monteporzio, Italy}
\date{Received {date will be inserted by the editor}; accepted {date
      will be inserted by the editor}}
  
  \abstract{We describe the advantages of using Integral Field
    Spectroscopy to observe deep fields of galaxy. The VIMOS Integral
    Field Unit is particularly suitable for this kind of studies
    thanks to its large field-of-view ($\sim 1$ arcmin$^{2}$).  After
    a short description of the VIMOS-IFU data reduction, we detail the
    main scientific issues which can be addressed using observations
    of the Hubble Deep Field South with a combination of Integral
    Field Spectroscopy and broad band optical and Near-Infrared
    imaging.
     }
\correspondence{foucaud@mi.iasf.cnr.it}

\maketitle

\section{Introduction}
\label{sec:intro}

Studying high redshift galaxies is one of the main topics that will
help in constructing a coherent picture of the physical processes that
led to galaxy evolution. 

Within the framework of large-scale structure formation, in which
small fluctuations of matter density grow under the influence of
gravity to form large-scale structures and galaxy halos, physically
motivated prescriptions have been used to describe the main processes
involved in galaxy formation and evolution (e.g.
\citeauthor{1991ApJ...379...52W},~\citeyear{1991ApJ...379...52W};
\citeauthor{2000MNRAS.319..168C},~\citeyear{2000MNRAS.319..168C}).
Furthermore perturbation theory and numerical simulations provide
useful predictions which can be compared with observations.  However,
comparing these models to available observations has not been
straightforward, and any discrepancy between the predictions and
observations could have important impacts on our understanding of some
fundamental processes. It is therefore important to perform those
comparisons at intermediate and high redshift, as here is where the
physical processes of galaxies formation are more visible.

In this paper, we report on the advantages of using Integral Field
Spectroscopy (IFS) associated to broad band imaging (optical and near
infrared) to observe deep fields compared to classical photometric
and spectroscopic observations. It is organised as follows: in
section~\ref{sec:photspec}, we shortly raise the limitations of the
photometric and spectroscopic surveys; in section~\ref{sec:3d} we
describe the advantages of using IFS; in section~\ref{sec:VIMOS} we
describe the VIMOS Integral Field Unit (IFU) and detail the data reduction
process; in section~\ref{sec:hdfs} we describe the use of VIMOS-IFU to
observe the Hubble Deep Field South and the main issues which will
follow from those observations. Finally, conclusions are summarised in
section~\ref{sec:sum}.

\section{Photometric and spectroscopic surveys}
\label{sec:photspec}

As a statistically large number of intermediate and high redshift
galaxies is needed to perform comparison between models and
observations, some large galaxy surveys were accomplished during
the last decade or are under acquisition (e.g.
CFRS~--~\citeauthor{1995ApJ...455...50L},~\citeyear{1995ApJ...455...50L};
CNOC~--~\citeauthor{2000ApJ...542...57C},~\citeyear{2000ApJ...542...57C};
DEEP~--~\citeauthor{2003SPIE.4834..161D},~\citeyear{2003SPIE.4834..161D};
VIRMOS~--~\citeauthor{2003Msngr.111...18L},~\citeyear{2003Msngr.111...18L}).

Two complementary kind of surveys exists: the deep multi-colour
photometric surveys, such as the VIRMOS Deep Imaging Survey (VDIS --
\citeauthor{vdis1},~\citeyear{vdis1} ;
\citeauthor{vdis2},~\citeyear{vdis2}) -- 4 fields of 4 deg$^2$ covered
in $BVRI$ bands and a sub-fraction in $U$ and $JK'$ -- and the
spectroscopic surveys, such as the VIRMOS-VLT Deep Survey (VVDS --
\citeauthor{2003Msngr.111...18L},~\citeyear{2003Msngr.111...18L}) --
around 150,000 redshifts acquired on the 4 fields of 4deg$^2$ of the
VDIS using the VIMOS spectrograph at VLT.  Photometric surveys allow
to access to a very large amount of objects (more than 2,000,000 for
$I_{AB}<25$ in the case of the VDIS) but with colour informations only
($UBVRI$, plus $JK'$ for a small fraction, in the case of the VDIS).
Those photometric surveys can then be used as reference catalogues for
spectroscopic surveys, but even with the new generation of
Multi-Object Spectrograph (like VIMOS) only a fraction of the
objects can have a spectroscopic follow-up (in the case of the VVDS
around 150,000 spectra are acquired).

A way to have all redshifts for those galaxies is to use photometric
methods, as the photometric redshifts technique (e.g.
\citeauthor{2000AJ....120.2206F},~\citeyear{2000AJ....120.2206F};
\citeauthor{2000A&A...363..476B},~\citeyear{2000A&A...363..476B};
\citeauthor{2002MNRAS.329..355A},~\citeyear{2002MNRAS.329..355A}) or
the Lyman-break technique (e.g.
\citeauthor{1999ApJ...519....1S},~\citeyear{1999ApJ...519....1S};
\citeauthor{2001ApJ...558L..83O},~\citeyear{2001ApJ...558L..83O};
\citeauthor{cfdf2},~\citeyear{cfdf2}) which are using the multi-colour
information to estimate the redshift and the nature of the galaxies.
Those techniques can be considered as very low resolution
spectroscopy.

To summarise, the large photometric and spectroscopic surveys allow to
gather large galaxy samples, but in one case with a poor spectral
information and in the other case a poor spatial sampling.

\section{Advantages and disadvantages of the 3D spectroscopy}
\label{sec:3d}

Thanks to Integral Field Spectroscopy (IFS), it is now possible to
gain both spectroscopic and photometric information in a given field
of view. The IFS is based on a new kind of instrument (Integral Field
Unit -- IFU) that provide a spectrum for each spatial element (spaxel)
thanks to optical fibres and/or micro-lenses and a dispersive element
(e.g.
\citeauthor{1995A&AS..113..347B},~\citeyear{1995A&AS..113..347B}).

The IFU configuration presents a lot of advantages for observing high
redshift galaxies. Their identification is free from any selection
criteria, such as the a priori selection applied in the photometric
and spectroscopic surveys. It does not impose any particular geometry
on object sampling and allows to extract spectra over the full object
extension, avoiding the slit loss problem faced with conventional
spectrographs. Finally IFUs are particularly sensitive to faint
slightly extended objects, like low surface brightness galaxies which
are difficult to detect and observe with conventional photometry and
spectroscopy.

Current generation IFUs are still only able to cover a small field of
view, their area coverage being below $1$~arcmin$^2$ (c.f.
section~\ref{subsec:desc-ifu}). In order to increase the field of
view, it is possible mosaicing multiple pointings, but it is still
unconceivable to cover areas of the same order as photometric and
spectroscopic surveys. Another disadvantage of these intruments is its
smaller total efficiency compared to ordinary spectroscopy, which
implies a longer integration time to reach the same depth of the
ordinary spectroscopic surveys.  Anyway these instruments are
particularly suited to observe small deep pointings like for instance
the Hubble Deep Fields (c.f. section~\ref{sec:hdfs}), even if it is
still to be demonstrated that IFUs can produce scientifically relevant
results for cosmology and high redshift galaxies.

\section{VIMOS Integral Field Unit}
\label{sec:VIMOS}

\subsection{Description of the VIMOS-IFU}
\label{subsec:desc-ifu}

\begin{table*}[t]
\begin{center}
\begin{tabular}{p{0.001cm}p{0.001cm}p{0.4cm}l*{5}{c}}
\multicolumn{4}{c}{Spectral resolution} & Wavelengths & Field &  Spatial resolution &
Spatial elements & Spectral elements\\ 
 & & & & ($\mu$m) & (arcsec$^2$) & (arcsec) & & \\
\hline
 R & = & 250 & (low) & $0.37-1$ & $54 \times 54$ & $0.67$ & 6400 & 600 \\
   &   &     &       &          & $27 \times 27$ & $0.33$ &      &     \\
 R & = & 2500 & (high) & $0.37-1$ & $27 \times 27$ & $0.67$ & 1600 & 4096 \\
   &   &      &        &          & $13 \times 13$ & $0.33$ &      &  \\
\end{tabular}
\caption{Characteristics of the VIMOS Integral Field Unit. The
  wavelength coverage is obtained thanks to two type of grisms: a red
  ($\sim 0.6-1\mu$m) and a blue ($\sim 0.37-0.7\mu$m). \vspace{-.5cm}}
\label{tab:VIMOS-IFU}
\end{center}
\end{table*}

The VIMOS instrument installed at the VLT-Paranal telescope U3 works
with two configurations: a spectro-imaging multi-object mode and an
integral field mode \citep{2003SPIE.4841.1670L}. The four
spectrographs that compose VIMOS are in common to those two different
modes. Each spectrograph is a classical focal reducer imaging
spectrograph, with a collimator, a parallel beam where dispersive
grisms are inserted, followed by a camera focusing onto a $2048 \times
4096$ $15\mu$m pixels CCD that produce an image for one quadrant. The
Mask Support Unit placed at the entrance focal plane of VIMOS allows
to position either four Invar masks in the case of the multi-object
mode or the four masks of the IFU and then switch from one mode to the
other \citep{2003SPIE.4841.1771B}.

The field of the IFU is sampled thanks to a 6400 fibres/lenslets
matrix. The output of the fibres are distributed on the four
spectrographs through the masks and coupled to each spectrograph
thanks to micro-lenses and curved prisms.  Two possible spatial
resolutions are available thanks to a removable focal elongator placed
in front of the IFU head. To obtain high spectral resolution, a
motorised shutter can be used to select only the central part of the
field in purpose to avoid spectra superposition. The characteristics
of the VIMOS-IFU are summarised in table~\ref{tab:VIMOS-IFU}.

\subsection{Reduction of VIMOS-IFU data}
\label{subsec:data}

The huge amount of data (6400 spectra from each exposure) provided by
VIMOS-IFU and the complexity of IFU data, require a fully dedicated
automatic data reduction pipeline. In this section we will quickly
describe the different steps of the data reduction (part of the data
reduction is also developed in
\citeauthor{2001adass..10..451S},~\citeyear{2001adass..10..451S}).

As 4 images are acquired during each VIMOS exposure (one image for
each quadrant), the data reduction is done on single quadrants up to
Data Cube reconstruction.  

For each frame the first usual steps of bias and flat field
corrections are applied.  Using those corrected frames the next step
is to extract the 2D spectra.  As the gap between the fibres is known,
this extraction is accomplished automatically starting from a first
guess. Wavelength calibration is then applied on each individual 2D
spectrum . 1D spectra are then obtained from the 2D spectra by adding
together the flux collected by the corresponding optical fibres.  As
the different fibres have different transmission efficiencies, a fibre
relative transmission correction is computed using the spectrum
continuum and sky lines. A standard way to compute this correction is
to first fit the continuum (by a 2 degrees polynomium) either on a
spectrum of twighlight sky or better on an image obtained adding the
set of images taken within a jitter offset sequence. A further finer
calibration can be done after the 1D extraction this time fitting the
sky lines (by a gaussian). From those fitted lines their flux is
determined and a refined relative normalisation is computed.  Then the
1D spectra are to be sky subtracted. To build an estimate of the sky
background intensity, it is necessary to identify the pure sky spectra
from the whole spectra set. Using the distribution of total light
intensities registered in the various spectra, pure sky spectra are
identified as those that have an intensity around the mode of the
distribution. Sky spectra are median-averaged together and this
spectrum is subtracted from all the 1D spectra.

In the case of several exposure per pointing, jitter offsets are
applied between exposures in order to better determine the relative
transmission (see above) and ameliorate the combination. Those offsets
between exposures are typically of around 4 fibres, i.e.  $\sim
2.7$~arcsec.. Thanks to jitter sequences it is also possible to remove
part of the fringing present in the red part of the spectra. This
fringing is due to reflection of the sky emission inside the detector
and a sigma-clipping medianing combination of the exposures, without
having previously shifted them, helps to eliminate this effect.

The combinations of the final spectra are obtained by medianing the
spectra from the various frames. Usage of the median instead of the
mean helps in the case of over-correction for fibre transmission or of
sky over-subtraction, as well as with cosmics removal.

To reconstruct a data cube from the 1D spectrum, the correspondence
between fibre position on IFU head -- which, coupled with world
coordinate system, gives fibre position on sky -- and spectrum
position on detection head is stored in a so-called ``IFU table''.
After reduction procedures, this table contains also informations
about fibre relative transmission measured by calibration procedures
and the fibre profile parameters (X and Y FWHM of the spectrum on the
CCD).  When all the four quadrants have been reduced, the fully
calibrated 1D spectra are rearranged in a data cube according to the
IFU table, to allow a spatially coherent reconstruction of the
observed sky region.

From the final data cube, 2D reconstructed images can be re-extracted
by collapsing the data cube in the wavelength direction using the
whole grism spectral range or an user-selected range. Interpolation or
drizzling techniques can then be used to better display the 2D
reconstructed images.

\section{Observing the Hubble Deep Field South}
\label{sec:hdfs}

The Hubble Deep Field South (HDF-S) is one of the best field observed
up-to-date in term of depth, multicolour extension and high spatial
resolution. This field was originally observed using the HST-WFPC2
camera in four wide bandpasses ($F300W$, $F450W$, $F606W$ and $F814W$)
\citep{2000AJ....120.2747C}. Very deep Near-Infrared ISAAC images are
also available in three wide bandpasses ($J$, $H$ and $K$), and a
complete multicolour catalogue of galaxies is selected down to
$K_{AB}=25$ \citep{2003ApJ...594L...9F,2001AJ....122.2190V}.  A
spectroscopic analysis of faint galaxies in that field was done using
the VLT-FORS spectrograph \citep{2002A&A...396..847V} but it is
limited by the multiplexing capabilities of the spectrograph.
Therefore a deeper spectroscopic sample of galaxies in that field is
needed.

Given the small size of the HDF-S field ($\sim 4.9$ arcmin$^2$), the
use of the VIMOS-IFU is perfectly suitable as only 7 IFU pointings are
requested to cover that field. Such observations will provide for the
first time the full spectroscopic coverage of galaxies with
$I_{AB}<26$ of the HDF-S (PI E. Giallongo).  A total time of 120h have
been allocated by ESO to observe this field using the VIMOS-IFU --
half of it being already allocated this year. For one pointing the
total observation time requested in low resolution mode is of 20h with
the red grism, i.e. a jitter sequence with exposures of 40min each,
and of 13h with the blue grism, i.e. a jitter sequence with exposures
of 40min each. Indeed, an exposure time of $T_{exp}=12$h is needed
with the red grism in low resolution mode to reach at $I_{AB}=26$ a
$S/N=4\,(8)$ per spectral resolution element at
$\lambda=7800\,(5800)$\AA$\,$ for point sources, and after integration
over all the fibers covering the source.  With the blue grism in low
resolution mode, the same $S/N$ is obtained with an exposure time of
$T_{exp}=8$h at $\lambda=3700\,(5500)$\AA$\,$.

Thanks to this new deep sample associated with the deep Near-Infrared
photometry, a better assessment of the star light distribution in a
wavelength range where it is little affected by dust absorption will
be possible.

The detection and confirmation of very high redshift galaxies
($z=5-6$) should be one of the first issues. Several recent studies
have shown that high redshift Ly-$\alpha$ emitters can be selected
thanks to specially dedicated narrow and broad band filters (e.g.
\citeauthor{2002ApJ...568L..75H},~\citeyear{2002ApJ...568L..75H};
\citeauthor{2003A&A...405L..19C},~\citeyear{2003A&A...405L..19C}). But
some galaxies, with strong emission line and a continuum too faint for
colour selection, can only be detected thanks to IFUs. Including the
contribution of such objects, a better estimation of the cosmologic
evolution of average star formation rate at high redshift should be
obtained.

Some recent results point towards the presence in the range of
redshift $z=2-3$ of an excess of bright star forming galaxies in the
rest frame blue luminosity function -- excess with respect to CDM
models predictions (see e.g.
\citeauthor{2003ApJ...593L...1P},~\citeyear{2003ApJ...593L...1P}).
The presence of such an excess could be assessed by the spectroscopic
confirmation of galaxies with $K_{AB}<25$ at $z=2-3$. This study
should bring strong informations on the star formation activity and
relative age of the bright high redshift galaxies. The detailed study
of morphology of those galaxies in rest frame Ultraviolet bands (HST
optical bands) and in rest frame optical bands (ISAAC NIR bands) could
also provide informations about spatial variation in the star
population and relative fraction of high redshift galaxies with red
nuclei.

Using VIMOS blue grism, redshift and spectral properties for galaxies
with $z=1-2$ will also be accessible through detection of interstellar
lines (for instance the line $CIV[1549$\AA$]$ at $z>1.4$ is shifted to
$\lambda>3700$\AA). This intermediate range of redshift is
particularly interesting as some studies indicate the star formation
rate tend to be at its maximum in that range
\citep{1996MNRAS.283.1388M,1997ApJ...486L..11C}.  The spatial
resolution reached by the VIMOS-IFU would also be able to probe
emission and absorption properties of substructures in bright $z=1-2$
galaxies.

Thanks to observations in multicolour broad band imaging extended to
the Near-Infrared wavelengths and comparisons with spectral synthesis
model, the galactic stellar mass can be estimated for a wide redshift
range in a given sample \citep{2000AJ....120.2206F}. This technique
allows to gather physical informations on the properties of high
redshift galaxies.  Furthermore main physical quantities of each
galaxy in the sample can also be estimated thanks to that technique,
as for instance the age of the last major starburst, the stellar mass
or the dust content. For instance, \cite{2003ApJ...594L...9F} were
able to constraint the stellar masses of galaxies in the HDF-S to
within of factor of two with $U-$ to $K-$ band imaging over a wide
redshift range ($0.5<z<3$). This study suffer anyway of a lack of
spectroscopic redshifts, which would help in decreasing the
uncertainties on the mass estimation. As a depth of $I_{AB}\sim26$ is
reached at a $S/N\sim 5$ with $\sim10$h of integration time using
VIMOS-IFU, and as according to \cite{2001AJ....122.2190V} the mean
color of the galaxies is $(I-K)_{AB}\sim 1$ when the Extremely Red
Objects are excluded, the IFU survey will gather a complete
spectroscopic sample and allow to determine the stellar mass for
Near-Infrared galaxies down to $K_{AB}=25$. Of course the accuracy of
this method is not comparable to kinematical observations but it is
enough to study the relative evolution of the stellar mass function at
different redshifts.

\section{Summary}
\label{sec:sum}

In this paper, we have described how suitable is the use of the VIMOS
Integral Field Unit to observe small deep fields, like the Hubble Deep
Field South. IFU observations allow to bypass the limitations of
conventional spectrographs, like mainly their multiplexing
capabilities. Furthermore the large size of the VIMOS-IFU is
particularly suitable for deep field observations, as only mosaic of
few pointings is required.

After a short description of the data reduction, we detailed the main
scientific issues which could be done thanks to observations of the
Hubble Deep Field South with IFU and broad band imaging extended to
the Near-Infrared wavelengths. For instance, those observations should
lead to the detection of high redshift galaxies ($z=5-6$), help to
confirm the excess of star forming galaxies in the redshift range
$z=2-3$, understand some of the spectral properties of galaxies in the
redshift range $z=1-2$ and allow an analysis of stellar mass
distribution for faint Near-Infrared selected galaxies.  All those
studies will bring important informations to better understand the
physical processes which are underlying to the galaxy evolution.

\vspace{.1cm}
\acknowledgements
This work has been performed under the framework of the VIRMOS consortium.
SF's work has been supported by RTN Euro3D postdoctoral fellowship.

\bibliographystyle{aa}

\begin{thebibliography}{28}
\expandafter\ifx\csname natexlab\endcsname\relax\def\natexlab#1{#1}\fi

\bibitem[{{Arnouts} {et~al.}(2002){Arnouts}, {Moscardini}, {Vanzella},
  {Colombi}, {Cristiani}, {Fontana}, {Giallongo}, {Matarrese}, \&
  {Saracco}}]{2002MNRAS.329..355A}
{Arnouts}, S., {Moscardini}, L., {Vanzella}, E., {et~al.} 2002, MNRAS, 329,
  355+

\bibitem[{{Bacon} {et~al.}(1995){Bacon}, {Adam}, {Baranne}, {Courtes}, {Dubet},
  {Dubois}, {Emsellem}, {Ferruit}, {Georgelin}, {Monnet}, {Pecontal},
  {Rousset}, \& {Say}}]{1995A&AS..113..347B}
{Bacon}, R., {Adam}, G., {Baranne}, A., {et~al.} 1995, A\&AS, 113, 347

\bibitem[{{Bolzonella} {et~al.}(2000){Bolzonella}, {Miralles}, \&
  {Pell{\'o}}}]{2000A&A...363..476B}
{Bolzonella}, M., {Miralles}, J.-M., \& {Pell{\'o}}, R. 2000, A\&A, 363, 476

\bibitem[{{Bonneville} {et~al.}(2003){Bonneville}, {Pri\'eto}, {Le~F\`evre},
  {Saisse}, {Voet}, {Zanichelli}, {Garilli}, {Vettolani}, {Maccagni},
  {Mancini}, \& {Picat}}]{2003SPIE.4841.1771B}
{Bonneville}, C., {Pri\'eto}, E., {Le~F\`evre}, O., {et~al.} 2003, in
  Instrument Design and Performance for Optical/Infrared Ground-based
  Telescopes. Edited by Iye, Masanori; Moorwood, Alan F. M. Proceedings of the
  SPIE, Volume 4841, pp. 1771-1782 (2003)., 1771--1782

\bibitem[{{Carlberg} {et~al.}(2000){Carlberg}, {Yee}, {Morris}, {Lin}, {Hall},
  {Patton}, {Sawicki}, \& {Shepherd}}]{2000ApJ...542...57C}
{Carlberg}, R.~G., {Yee}, H. K.~C., {Morris}, S.~L., {et~al.} 2000, ApJ, 542,
  57

\bibitem[{{Casertano} {et~al.}(2000){Casertano}, {de~Mello}, {Dickinson},
  {Ferguson}, {Fruchter}, {Gonzalez-Lopezlira}, {Heyer}, {Hook}, {Levay},
  {Lucas}, {Mack}, {Makidon}, {Mutchler}, {Smith}, {Stiavelli}, {Wiggs}, \&
  {Williams}}]{2000AJ....120.2747C}
{Casertano}, S., {de~Mello}, D.~F., {Dickinson}, M., {et~al.} 2000, AJ, 120,
  2747

\bibitem[{{Cole} {et~al.}(2000){Cole}, {Lacey}, {Baugh}, \&
  {Frenk}}]{2000MNRAS.319..168C}
{Cole}, S., {Lacey}, C.~G., {Baugh}, C.~M., \& {Frenk}, C.~S. 2000, MNRAS, 319,
  168

\bibitem[{{Connolly} {et~al.}(1997){Connolly}, {Szalay}, {Dickinson},
  {Subbarao}, \& {Brunner}}]{1997ApJ...486L..11C}
{Connolly}, A.~J., {Szalay}, A.~S., {Dickinson}, M., {Subbarao}, M.~U., \&
  {Brunner}, R.~J. 1997, ApJ, 486, L11+

\bibitem[{{Cuby} {et~al.}(2003){Cuby}, {Le F{\` e}vre}, {McCracken},
  {Cuillandre}, {Magnier}, \& {Meneux}}]{2003A&A...405L..19C}
{Cuby}, J.-G., {Le F{\` e}vre}, O., {McCracken}, H.~J., {et~al.} 2003, A\&A,
  405, L19

\bibitem[{{Davis} {et~al.}(2003){Davis}, {Faber}, {Newman}, {Phillips},
  {Ellis}, {Steidel}, {Conselice}, {Coil}, {Finkbeiner}, {Koo}, {Guhathakurta},
  {Weiner}, {Schiavon}, {Willmer}, {Kaiser}, {Luppino}, {Wirth}, {Connolly},
  {Eisenhardt}, {Cooper}, \& {Gerke}}]{2003SPIE.4834..161D}
{Davis}, M., {Faber}, S.~M., {Newman}, J., {et~al.} 2003, in Discoveries and
  Research Prospects from 6- to 10-Meter-Class Telescopes II. Edited by
  Guhathakurta, Puragra. Proceedings of the SPIE, Volume 4834, pp. 161-172
  (2003)., 161--172

\bibitem[{{Fontana} {et~al.}(2000){Fontana}, {D'Odorico}, {Poli}, {Giallongo},
  {Arnouts}, {Cristiani}, {Moorwood}, \& {Saracco}}]{2000AJ....120.2206F}
{Fontana}, A., {D'Odorico}, S., {Poli}, F., {et~al.} 2000, AJ, 120, 2206

\bibitem[{{Fontana} {et~al.}(2003){Fontana}, {Donnarumma}, {Vanzella},
  {Giallongo}, {Menci}, {Nonino}, {Saracco}, {Cristiani}, {D'Odorico}, \&
  {Poli}}]{2003ApJ...594L...9F}
{Fontana}, A., {Donnarumma}, I., {Vanzella}, E., {et~al.} 2003, ApJ, 594, L9

\bibitem[{{Foucaud} {et~al.}(2003){Foucaud}, {McCracken}, {Le~F\`evre},
  {Brodwin}, {Arnouts}, {Lilly}, {Crampton}, \& {Mellier}}]{cfdf2}
{Foucaud}, S., {McCracken}, H.~J., {Le~F\`evre}, O., {et~al.} 2003,
  A\&A, 409, 835

\bibitem[{{Hu} {et~al.}(2002){Hu}, {Cowie}, {McMahon}, {Capak}, {Iwamuro},
  {Kneib}, {Maihara}, \& {Motohara}}]{2002ApJ...568L..75H}
{Hu}, E.~M., {Cowie}, L.~L., {McMahon}, R.~G., {et~al.} 2002, ApJ, 568, L75

\bibitem[{{Le~F{\` e}vre} {et~al.}(2003{\natexlab{a}}){Le~F{\` e}vre},
    {Vettolani}, {Maccagni}, {Picat}, {Garilli}, {Tresse}, {Adami},
    {Arnaboldi}, {Arnouts}, {Bardelli}, {Bolzonella}, {Bottini},
    {Buzzarello}, {Charlot}, {Chincarini}, {Contini}, {Foucaud},
    {Franzetti}, {Guzzo}, {Gwyn}, {Ilbert}, {Iovino}, {Le Brun},
    {Longhetti}, {Marinoni}, {Methez}, {Mazure}, {McCracken},
    {Mellier}, {Meneux}, {Merluzzi}, {Paltani}, {Pell{\` o}}, {Pollo},
    {Radovich}, {Rippepi}, {Rizzo}, {Scaramella}, {Scodeggio},
    {Zamorani}, {Zanichelli}, \& {Zucca}}]{2003Msngr.111...18L}
  {Le~F{\` e}vre}, O., {Vettolani}, G., {Maccagni}, D., {et~al.}
  2003{\natexlab{a}}, The Messenger, 111, 18

\bibitem[{{Le~F\`evre} {et~al.}(2003{\natexlab{b}}){Le~F\`evre}, {Mellier},
  {McCracken}, {Foucaud}, {Gwyn}, {Radovich}, {Dantel-Fort}, {Bertin},
  {Moreau}, {Cuillandre}, {Pierre}, {Le~Brun}, {Mazure}, \& {Tresse}}]{vdis1}
{Le~F\`evre}, O., {Mellier}, Y., {McCracken}, H.~J., {et~al.}
  2003{\natexlab{b}}, {\it A\&A, accepted} -- astro-ph/0306252

\bibitem[{{Le~F\`evre} {et~al.}(2003{\natexlab{c}}){Le~F\`evre}, {Saisse},
  {Mancini}, {Brau-Nogue}, {Caputi}, {Castinel}, {D'Odorico}, {Garilli},
  {Kissler-Patig}, {Lucuix}, {Mancini}, {Pauget}, {Sciarretta}, {Scodeggio},
  {Tresse}, \& {Vettolani}}]{2003SPIE.4841.1670L}
{Le~F\`evre}, O., {Saisse}, M., {Mancini}, D., {et~al.} 2003{\natexlab{c}}, in
  Instrument Design and Performance for Optical/Infrared Ground-based
  Telescopes. Edited by Iye, Masanori; Moorwood, Alan F. M. Proceedings of the
  SPIE, Volume 4841, pp. 1670-1681 (2003)., 1670--1681

\bibitem[{{Lilly} {et~al.}(1995){Lilly}, {Le~F\`evre}, {Crampton}, {Hammer}, \&
  {Tresse}}]{1995ApJ...455...50L}
{Lilly}, S.~J., {Le~F\`evre}, O., {Crampton}, D., {Hammer}, F., \& {Tresse}, L.
  1995, ApJ, 455, 50

\bibitem[{{Madau} {et~al.}(1996){Madau}, {Ferguson}, {Dickinson}, {Giavalisco},
  {Steidel}, \& {Fruchter}}]{1996MNRAS.283.1388M}
{Madau}, P., {Ferguson}, H.~C., {Dickinson}, M.~E., {et~al.} 1996, MNRAS, 283,
  1388

\bibitem[{{McCracken} {et~al.}(2003){McCracken}, {Radovich}, {Bertin},
  {Mellier}, {Dantel-Fort}, {Le~F\`evre}, {Cuillandre}, {Gwyn}, {Foucaud}, \&
  {Zamorani}}]{vdis2}
{McCracken}, H.~J., {Radovich}, M., {Bertin}, E., {et~al.} 2003, A\&A,
  410, 17

\bibitem[{{Ouchi} {et~al.}(2001){Ouchi}, {Shimasaku}, {Okamura}, {Doi},
  {Furusawa}, {Hamabe}, {Kimura}, {Komiyama}, {Miyazaki}, {Miyazaki}, {Nakata},
  {Sekiguchi}, {Yagi}, \& {Yasuda}}]{2001ApJ...558L..83O}
{Ouchi}, M., {Shimasaku}, K., {Okamura}, S., {et~al.} 2001, ApJ, 558, L83

\bibitem[{{Poli} {et~al.}(2003){Poli}, {Giallongo}, {Fontana}, {Menci},
  {Zamorani}, {Nonino}, {Saracco}, {Vanzella}, {Donnarumma}, {Salimbeni},
  {Cimatti}, {Cristiani}, {Daddi}, {D'Odorico}, {Mignoli}, {Pozzetti}, \&
  {Renzini}}]{2003ApJ...593L...1P}
{Poli}, F., {Giallongo}, E., {Fontana}, A., {et~al.} 2003, ApJ, 593, L1

\bibitem[{{Scodeggio} {et~al.}(2001){Scodeggio}, {Zanichelli}, {Garilli},
  {Le~F{\` e}vre}, \& {Vettolani}}]{2001adass..10..451S}
{Scodeggio}, M., {Zanichelli}, A., {Garilli}, B., {Le~F{\` e}vre}, O., \&
  {Vettolani}, G. 2001, in ASP Conf. Ser. 238: Astronomical Data Analysis
  Software and Systems X, 451--+

\bibitem[{{Steidel} {et~al.}(1999){Steidel}, {Adelberger}, {Giavalisco},
  {Dickinson}, \& {Pettini}}]{1999ApJ...519....1S}
{Steidel}, C.~C., {Adelberger}, K.~L., {Giavalisco}, M., {Dickinson}, M., \&
  {Pettini}, M. 1999, ApJ, 519, 1

\bibitem[{{Vanzella} {et~al.}(2002){Vanzella}, {Cristiani}, {Arnouts},
  {Dennefeld}, {Fontana}, {Grazian}, {Nonino}, {Petitjean}, \&
  {Saracco}}]{2002A&A...396..847V}
{Vanzella}, E., {Cristiani}, S., {Arnouts}, S., {et~al.} 2002, A\&A, 396, 847

\bibitem[{{Vanzella} {et~al.}(2001){Vanzella}, {Cristiani}, {Saracco},
  {Arnouts}, {Bianchi}, {D'Odorico}, {Fontana}, {Giallongo}, \&
  {Grazian}}]{2001AJ....122.2190V}
{Vanzella}, E., {Cristiani}, S., {Saracco}, P., {et~al.} 2001, AJ, 122, 2190

\bibitem[{{White} \& {Frenk}(1991)}]{1991ApJ...379...52W}
{White}, S. D.~M. \& {Frenk}, C.~S. 1991, ApJ, 379, 52

\end{thebibliography}

\end{document}